# Generic Rolling Access to Synchrotron Radiation Facilities

*Arka Bikash Dey\*, Kai Bagschik, Ruslan Khubbutdinov, Jan-Peter Kurz, Daniela Unger, Florian Bertram, Johanna Hakanpää, Christoph Schlueter, Dmitri Novikov, Martin Tolkiehn, Milena Lippmann, Harald Reichert, Hans-Christian Wille, Oliver H. Seeck*

Deutsches Elektronen-Synchrotron DESY, Notkestraße 85, 22607 Hamburg, Germany

**A generic model for rolling access to synchrotron radiation experiments is presented, which has the capacity to replace call-based access models. Proposal submission, evaluation and scheduling are all executed in a rolling fashion. This significantly reduces the waiting times between proposal submission and experiment execution than that of the call-based access model. The generic rolling access model is in principle applicable to any beamlines, regardless of the number of experimental methods or setups it provides to users. This access model is flexible and could provide faster access compared to call-based access model as well as accommodate experiments and projects requiring extended preparation times.**

**Keywords:** synchrotron radiation, light source, access, generic rolling access, PETRA III, PETRA IV

Large-scale synchrotron radiation facilities [1-2] produce intense X-ray beams using high-energy electron storage rings. There are approximately 30 large-scale synchrotron radiation facilities around the world [3]. These facilities play a pivotal role in the advancement of numerous scientific disciplines [4-5] and commercial sectors [6], as they are capable of unraveling chemical compositions, molecular structures, and dynamic processes with unprecedented detail. The commercial sector benefits from these facilities drawing on support for, e.g., materials characterization, drug development, and advancements in manufacturing processes [6-8]. It is therefore crucial to enable generic user access to such cutting-edge facilities as this will accelerate discoveries and advancements across numerous fields, ultimately contributing to societal progress and long-term economic growth [9].

The majority of synchrotron radiation facilities operate under conventional (bi-annual or tri-annual call-based) access models since the 1990s, providing service to both scientific users and commercial customers. Academic users request access to X-ray beamtime with duration from a few hours (short-term projects) to up to few weeks (long-term projects) through submission of proposals. Following the submission of an academic proposal, it is subject to a review by an external expert panel held also bi- or tri-annually. The scheduling of X-ray beamtime is based on the rating for the academic proposals and the availability of beamtime. Therefore, an average wait time of approximately nine months and six months needed between the submission of an academic proposal and the actual beamtime for bi- or tri-annual call-based access, respectively. Commercial customers purchase beamtime through a contract and are receive priority scheduling. Although the conventional access models are well-established, they lack agility and responsiveness for urgent, collaborative, and long-term projects. In response to concerns raised by users and customers, adaptations such as rapid access, block allocation groups (BAG), and long-term projects (LTP), and several other proposal schemes [10] have been commonly implemented alongside the conventional access models. However, bi- or tri-annual proposal calls are still the main access route to beamtime. In addition, a few facilities, such as Spring-8 and Elettra, have adopted more frequent proposal calls per year, with six and twelve calls, respectively. Moreover, some synchrotron facilities like ESRF has adopted a rolling access system for some specific beamlines. These rolling and multiple calls are implemented for highly standardized and fully automated beamlines, such as beamlines for molecular or protein crystallography, and small angle X-ray scattering.

In addition, the conventional proposal-based beamtime access model lacks the flexibility to accommodate scientific projects that require access to multiple beamlines/beamtimes, complementary offline techniques, computing, data analysis, training, consulting, and certification. Alternatively, a project-based submission scheme could be introduced where the users can submit their projects with the required aforementioned demand. A combination of rolling access and project-based submission scheme would provide a higher degree of flexibility in coordinating and managing resources to ensure a higher rate of success for scientific research projects. A team of experts may coordinate all projects, manage access and optimize resource allocation. This will usher in a new era of accessibility and efficiency in research with synchrotron radiation.

Recognizing the need for change, as a first foot-step towards the generic rolling access-based project-proposal submission model, we introduce an alternative access model (generic rolling access) applicable to all types of synchrotron radiation experiments. This new procedure aims to replace the conventional access model, while meeting modern demands and preserving the strengths of the conventional access models. As a testing, generic rolling access-based proposal (not project-proposal) has been implemented at five beamlines at PETRA III, Deutsches Elektronen-Synchrotron DESY, Hamburg, Germany. In future, project-proposal submission could replace conventional proposal submission, and would be also included within this generic rolling access model. The included five beamlines can be categorized by different techniques as two scattering beamlines: the High-Resolution Diffraction Beamline (P08) [11-13] equipped with more than three different instrumental setups, the In-situ Diffraction and Imaging Beamline (P23) with one diffraction and one imaging end-station, one spectroscopy beamline: the Hard X-ray Photoelectron Spectroscopy Beamline (P22) [14-15] equipped with different ultra-high vacuum and ambient/high-pressure setups, and two crystallography beamlines: the Chemical Crystallography Beamline (P24) with two end-stations and the High-throughput Macromolecular Crystallographic Beamline (P11) [16-17]. In addition, access to preparation and characterization laboratories is included in the rolling access during this test phase at PETRA III through evaluation and scheduling in a rolling fashion as well. Based on an evaluation after the test phase, necessary refinements to this access model will be made. In the future, it is planned to extend this rolling procedure to other beamlines at PETRA III and eventually to all beamlines at the upcoming PETRA IV facility [18].

This new generic rolling access model offers five distinct advantages that are absent in most conventional access models; these are:

(a) a unified rolling project submission scheme,
(b) reduced waiting times between project proposal submission and experiment execution,
(c) higher reliability on timely project completion,
(d) enhanced flexibility in preparation time,
(e) and access to multiple beamlines and/or multiple beamtimes through a single project proposal.

However, this new rolling access model also presents a few significant challenges:
(i) establishing a transparent process for dynamic ranking in rolling reviews
(ii) ensuring scientific excellence of the individual projects,
(iii) handling project proposals with ranking close to the cut-off,
(iv) and implementing rolling scheduling for fast and flexible access.

In this article, we discuss, one by one, the advantages of the generic rolling access model in comparison to the conventional access, and the strategy taken to address the challenges in the rolling access model.

**1.1 A unified rolling proposal submission scheme**

Currently, almost all large-scale facilities offer a variety of proposal types, each with its own proposal template. In the conventional access models, users are notified of the opening of different calls, each with an announced deadline. In the rolling access model, project proposals can be submitted at any time throughout the year using a single project proposal submission scheme, eliminating the worry about call deadlines and a variety of submission schemes. This unified project submission scheme eliminates the necessity for separate calls for regular, LTP, BAG, and rapid access proposals. Furthermore, it reduces the number of proposal templates to just two, the first template (type-I) when requested shift is less than or equal to a certain number ($S^*$) and the second template (type-II) when the requested shift is above the certain number ($S^*$); as indicated in **Figure 1**. In the rolling access model, there is a maximum limit of ($S_{max}$) for the number of requested shifts per proposal. Projects requesting more than $S^*$ shifts (type II proposal as indicated in **Figure 1**) are required to provide a justification for such a large number of shifts, including methodological and/or strategic scientific goals and, if applicable, the associated benefits to other user groups of the beamline. This new strategy extends the strength of LTPs and BAGs to all proposals requested with more than $S^*$ shifts, making it more convenient and useful for some scientific projects with the need of long-term planning. During the testing phase at PETRA III, $S_{max}$ and $S^*$ are set to 72 shifts and 21 shifts, respectively. The introduction of this unified scheme removes the unnecessary complexities and deadlines in submitting proposals through different schemes, aiming at enhancing the satisfaction of and convenience to the researchers.

## 1.2 Reduced waiting times between project proposal submission and experiment execution

A significant proportion of users, particularly those in industry and academia with urgent needs, express dissatisfaction with the conventional user access procedure, which involves an average wait time of approximately nine months (bi-annual call-based access at PETRA III) between the submission of a proposal and the actual allocation of beamtime. This is in addition to the time spent formulating an idea and meeting the call deadline. The rolling procedure allows users to submit proposals at any time without any deadlines. Following the submission of a proposal, a technical and experimental feasibility check (also known as the internal review) will be conducted promptly as indicated in **Figure 1**. This is followed by the actual evaluation of academic proposals by external experts, which can take a few weeks depending on the evaluation procedure, which will be discussed later (see **Figure 2**). Immediately after review, proposals are passed into the scheduling process (see **Figure 3**). Depending on the complexity of the experiment and user requirements for experiment preparation, the first beamtime can be executed within a few weeks after the review or later if requested by the users. The scheduling strategies are discussed in a separate section (see **Figure 3**). This entire procedure significantly reduces the waiting time between proposal submission and experiment execution. In this new rolling access procedure, the average wait time from the project proposal submission to the earliest possible experimental time is of the order of two months for academic proposals with review, or two weeks for commercial customers at some beamlines. If the interval between the conceptualization of an idea and call deadline under the conventional access model is considered, we estimate that the rolling procedure could potentially reduce the total waiting time by a factor of three to five. This procedure is designed to meet the needs of urgent projects, industrial projects, rapid access, and priority access (through contract with cooperation partners) in this new access framework.

**1.3 Higher reliability in project completion**

The generic rolling access model aims to enhance the success rate of completing scientific projects associated with synchrotron research, thereby making it more attractive to both new and experienced synchrotron users. In the conventional user access procedure, an unsuccessful beamtime proposal can only be resubmitted after, for example, a period of six months due to the pre-fixed bi-annual calls. This results in a total waiting time stretching over a year and leaves the users with the problem of meeting project timelines including possible failure to reach the project's objectives. This is a particular concern for PhD students. The rolling procedure is designed to minimize such uncertainty by providing users with prompt feedback from experts and the potential of rapid re-submission after modification. This will attract users with new ideas, enabling them to include synchrotron experiments in their science projects with greatly reduced uncertainty in project timelines and objectives.

**1.4 Enhanced flexibility in preparation time**

The generic rolling procedure is structured in a manner that it not only supports rapid access projects but also allows for the flexibility to accommodate projects which require a longer preparation period. This is the actual time elapsed between the approval of the experiment and its execution. In the conventional access procedure, LTPs and BAGs are valid for a period of two years in general. In contrast, at many facilities, regular proposals are valid for one scheduling period, that is generally six months, only. Consequently, this restricts the maximum preparation time for any regular proposal. In the new rolling procedure, proposals may be valid for a longer period ($T_{validity}$) providing enhanced flexibility in terms of preparation time, which can extend beyond a scheduling period. This satisfies a wide range of needs for different types of projects, enabling experimental plans to be arranged according to the specific needs of each project. During the testing phase at PETRA III, $T_{validity}$ is set to be two years, which could, in addition, be optimized individually for each beamline.

## 1.5 Access to multiple beamlines and/or multiple beamtimes through a single project proposal

The distribution of the total number of approved shifts into multiple beamtime blocks is typical mostly for LTPs and BAGs, while regular proposals are, in general, associated with a single experimental session. This key feature of LTPs and BAGs is now open to all proposals (both type I and type II proposals, cf. **Figure 1**) within the framework of the generic rolling access procedure. The distribution of approved shifts over multiple beamtime blocks is a prerequisite for project-oriented science with milestones and deliverables. The option of distributing the allocated beamtime over the duration of a few years is particularly advantageous for users and customers of automated beamlines such as macromolecular crystallography beamlines, where a large amount of experimental data can be collected in a relatively short period of beamtime and the collected data need to be evaluated before proceeding with the next step. Therefore, splitting the assigned beamtime into multiple blocks significantly improves the flexibility in experiment strategy and quality of the data collection. Therefore, new ideas and sequential long-term projects can be easily incorporated through extending this advantage also to other disciplines through this new rolling procedure.

Similarly, access to multiple beamlines within a synchrotron facility is often necessary to successfully solve a research problem. However, most of the access models provide access to individual beamlines through reviewing the merits of individual proposals rather than the merits of the entire project. This can lead to a situation where users are only successful in obtaining partial access, resulting in partial or even complete failure of the project. Access to multiple beamlines through a single project proposal is therefore an important asset. In addition, need for consultation, training, complementary offline techniques, computing, and support for data analysis at the synchrotron source is expected to be grow with the advent of PETRA IV [18] and would be included under the project-based access model. Incorporating these into the new generic rolling access model will result in a higher success rate for projects, leading to the generation of more impactful scientific output and enabling faster progress in many fields of science in both academic and industrial research and development. At PETRA III, currently, users can access multiple beamtimes in a single beamline during the current testing phase with proposal-based access model. However, once the project-based access is included, it will also provide access to multiple beamlines through a single project proposal.

The realization of these five key features will result in a higher level of user satisfaction by meeting their various demands. Nevertheless, the implementation of this rolling procedure could pose challenges in the review process and in the subsequent scheduling process. These challenges are being addressed with proper strategies in the testing phase at PETRA III and will be further optimized based on the feedback obtained during this testing phase.

**1.6 Establishing a transparent process for dynamic ranking in rolling reviews**

In contrast to the conventional access models, where academic proposals are evaluated all together in every four to six months and ranked after the review process for beamtime allocation, the rolling review lacks direct comparability between the proposals, which are submitted at different times. To address this challenge, a new review procedure has been devised, which involves a combination of independent evaluations of each project proposal and holding regular meetings with the panel of experts for comparison with previously reviewed project proposals. The procedure is adapted as follows: After feasibility check, each project is first reviewed by the reviewers, and given an absolute rating (refer **Figure 2**). In the second step, which is necessary if the first step leaves uncertainties or if more than $T_{cutoff}$ shifts are requested, the new project proposal is ranked against previously rated project proposals (refer **Figure 2**) in a regular meeting within the review panel. The regular meetings will ensure a fair evaluation of the project proposals by resolving conflicts in ratings, and rankings. This regular meeting with the expert panel will also provide the forum to evaluate and discuss the merits of the project proposals and compare them with previously reviewed project proposals. Proposals with more than the S* number of shifts always require thorough discussion and are evaluated by the pool of expert reviewers in regular meetings. The time frame for the initial review is set to two weeks and a monthly meeting has been scheduled for review panel discussions during the test phase at PETRA III. The users are notified immediately on the results of the review, comments, suggestions for improvements and also an encouragement to resubmit improved project proposals where applicable.

**1.7 Ensuring scientific excellence of the individual projects**

The rated project proposals are ranked and sorted into a stack according to their ratings. Multiple proposal-stacks for one beamline could be set up in case there are several experimental techniques available at a single beamline that are not comparable. This will eliminate unfair competition between individual projects requesting completely different techniques. To ease scheduling, the beamline responsible prepares a public schedule of separate time-blocks reaching several weeks into the future for each setup or technique. The available shifts for a beamline can be divided into several blocks for such non-comparable techniques. This helps to avoid major loss in beamtime due to frequent changes of experimental setups at a beamline and conflicts between proposals requesting different techniques. Projects will be scheduled starting at the top of the stack according to the amount of beamtime available in the respective blocks. Commercial proposals will be given the highest priority and will always be placed at the top of the stack for immediate scheduling, as they do not undergo the external review process (see **Figure 2**). To ensure the quality of the accepted individual project proposals, the external panel experts may be invited to meet on a regular interval to discuss previously reviewed project proposals and to provide an overall assessment of the review process.

**1.8 Handling project proposals with ranking close to the cut-off**

The highest-rated project proposals will receive beamtime on demand while the low-rated proposals will receive a notification with the possibility for resubmission of the project proposals with changes suggested by the external experts in that field. However, uncertainties remain for project proposals receiving a decent rating close to the cut-off for receiving beamtime with little room for improvement. Such proposals might stay in the stack for an extended period of time (see **Figure 2**) without being scheduled. They would eventually be removed from the stack with a rejection notification, thus undermining the goal of providing a fast response. To prevent this, the probability for receiving beamtime can be calculated via comparison with the success rate of previous project proposals with equal rating. To facilitate this approach a record of all proposals with their individual wait times is kept for each beamline / experimental facility. After an initial time, i.e. once sufficient statistics are available, an accurate forecast can be provided to the users (principle investigator and project leader) of these projects immediately after the review, thus enabling them to take a timely decision on a potential retraction of the proposal and/or resubmission of a modified proposal.

### 1.9 Implementing rolling scheduling for fast and flexible access

A smart scheduling strategy is essential for making the generic rolling access a success. The scheduling strategy has to allow for different filling factors in the coming months to ensure the availability of slots for the incorporation of highly ranked urgent proposals, as well as the scheduling of highly ranked proposals with long preparation times well in advance. The coming months can be sliced into n periods, where each period $\Delta t_i$ ($i$ = 1, 2, 3, .., n) is associated with a separate filling factor $f_i$, where $f_i$ represents the ratio of the scheduled shifts to the number of available shifts within that period (see **Figure 3a**). For example, if there are slices with duration $\Delta t_1$, $\Delta t_2$, $\Delta t_3$, …, $\Delta t_n$ and corresponding filling factors $f_1$, $f_2$, $f_3$, …, $f_n$ then the first period ($\Delta t_1$) will be scheduled up to (100*$f_1$)%, the second period ($\Delta t_2$) up to (100*$f_2$)% and so on. Beyond the aggregated number of slices $T = \sum_i \Delta t_i$, no beamtime will be scheduled. As a starting point, a scheduling strategy based on three time periods can be adopted: $\Delta t_1$ for urgent proposals, $\Delta t_2$ for normal proposals, and $\Delta t_3$ for long-term requests, could be employed. Each of the three time periods can be filled with a specific filling factor $f_1$, $f_2$, and $f_3$ respectively. For instance, if $\Delta t_1$ = 1 month, $\Delta t_2$ = 2 months, $\Delta t_3$ = 3 months and $f_1$ = 1, $f_2$ = 0.3, and $f_3$ = 0.1, the period from now to 1 month in the future would be fully scheduled and the second period (months 2-3) would be filled only to 30%, the last period (months 4-6) only to 10% with experiments (see **Figure 3b**). From 7 months onwards, no experiments are scheduled. As an illustration, the parameter set ($\Delta t_1$, $\Delta t_2$, $\Delta t_3$, $f_1$, $f_2$, $f_3$, $c$, $u$) = (6 months, 0 months, 0 months, 1, 0, 0, 0 months, 6 months) resembles the current conventional access model at PETRA III, where 'c' is defined as the maximum time after which a proposal will be removed from the stack of proposal with a rejection notification to the users. 'u' defines how frequent the stack would be updated with new incoming proposals. In current conventional access model at PETRA, a 6-month period is fully scheduled, which is repeated every ($u =$) 6 months, and the proposal calls have a deadline, making this number 'c' is 0.

A simulation was conducted to assess the efficacy of a rolling access scheduling procedure. A total of 450 virtual beamtime requests were randomly distributed over 730 days of beamtime, with a random normally distributed final rating [set between 1.0 and 5.0; mean at rating 2.5 and standard deviation of 1.0], and with a random value for the requested beamtime shifts [set between 3 and 18 shifts] serving as input for the scheduling process. Here, 1 shift equals to 8 hours of beamtime, which is the case for PETRA III and most other synchrotrons. The two extreme points in the ratings, 1.0 and 5.0, indicate outstanding and poor quality, respectively. The ranking of the proposals is based on their rating: the better the quality of a proposal, the lower its ranking number, and vice versa. The ranking of the proposals is updated as soon as a new proposal enters the scheduling stack of proposals. Proposals are automatically removed from the stack after 3 months without scheduling as we set c = 3 months in the simulation. As an example, we present two simulation runs with a distinct set of parameters in each case. The values of the parameters in the simulation of the first run are as follows: ($\Delta t_1$, $\Delta t_2$, $\Delta t_3$, $f_1$, $f_2$, $f_3$, $c$, $u$) = (30 days, 60 days, 90 days, 1, 0.3, 0.1, 3 months, 1 week), referred to as case 1 (see **Figure 3c**), and in the second run (15 days, 85 days, 55 days, 1, 0.15, 0.12, 3 months, 1 week), referred to as case 2 (see **Figure 3d**), respectively. Figures 3b and 3c illustrate the results for the two case scenarios with correlation plots for the rank of the proposals in the proposal stack, waiting time, and experiment duration. In both cases the time period for the review is not accounted for. The duration of the experiment is encoded via the diameter of the circles, with the smallest circles corresponding to the allocation of 3 shifts and the largest circles corresponding to the allocation of 18 shifts. The circles featuring a black rim represent beamtimes that have been assigned one day less experimental time than requested. The colors of the circles do not convey any additional meaning; they are solely employed to enhance readability. The model simulation with different sets of parameters for this new scheduling scheme indicates that the top-ranked projects can be scheduled immediately, with an average wait time of two months between project proposal submission and actual beamtime for academic access. The scheduling strategy in case 2 is more aggressive than in case 1 as the value of the parameter $\Delta t_1$ is smaller for case 2 than it is for case 1. In case 1 only 2% of the available time could not be scheduled, and the higher-ranked proposals could be scheduled within less than 40 days. In contrast, for the more aggressive scheduling strategy (case 2), the higher-ranked proposals receive beamtime already 15 days after proposal submission (again excluding the time for the review). However, 3.3% of all days remain unscheduled in this case. Both simulations show that about 30% of the experimental sessions with longer duration have to be cut by one day or must allow for a distribution of the approved number of shifts into multiple beamtime blocks to enable adequate filling of the schedule. In the testing phase of the generic rolling access model at PETRA III, we start with the parameter values ($\Delta t_1$, $\Delta t_2$, $\Delta t_3$, $f_1$, $f_2$, $f_3$, $c$, $u$) = (2 months, 2 months, 2 months, 1, 0.7, 0.2, 6 months, 1 week). These parameters will be optimized individually for the five participating beamlines during the testing phase. A review of the rolling access scheme including the five beamline teams, safety and lab teams at DESY, the pool of reviewers, and the users will be held after the initial testing phase to assess the level of satisfaction in each stakeholder and to identify measures with a potential for improvement if necessary.

In summary, the generic rolling access procedure addresses long-standing challenges of the conventional access model while maintaining the positive aspect of the conventional access model and addressing diverse user needs. Through its seamless and convenient project proposal submission process, the rolling procedure minimizes waiting times, and facilitates rapid response to urgent and long-term research demands. The potential challenges in the review and scheduling process have been addressed in our proposed scheduling schemes. Simulations show promising results, which are being refined via a testing phase for the rolling access procedure at PETRA III. Following refinement, the generic rolling access model has, in conjunction with the inclusion of extended offline supporting techniques, the potential to revolutionize the user access model in any large-scale synchrotron radiation facility in the years to come.

Figures:

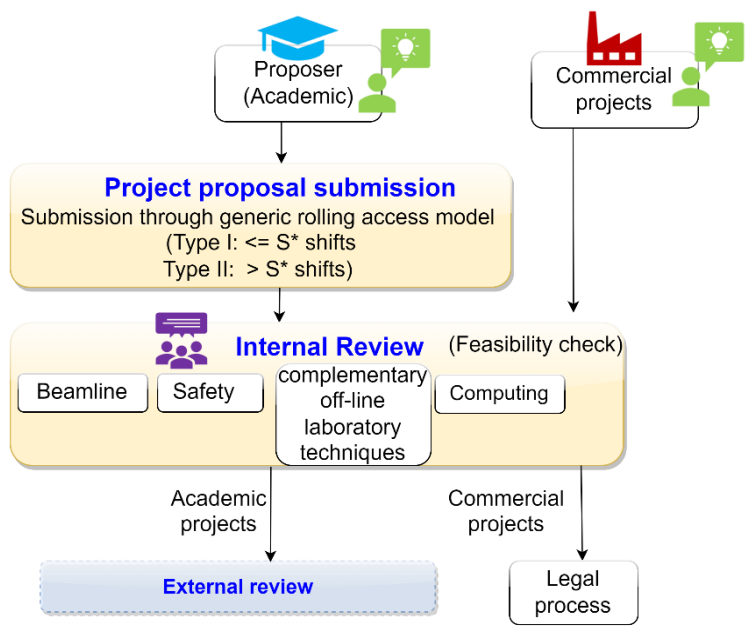

Figure 1. Workflow illustrating the unified proposal submission scheme for the new rolling access model. This includes the use of common proposal templates and simultaneous feasibility checks. The detailed external review process is shown in Figure 2.

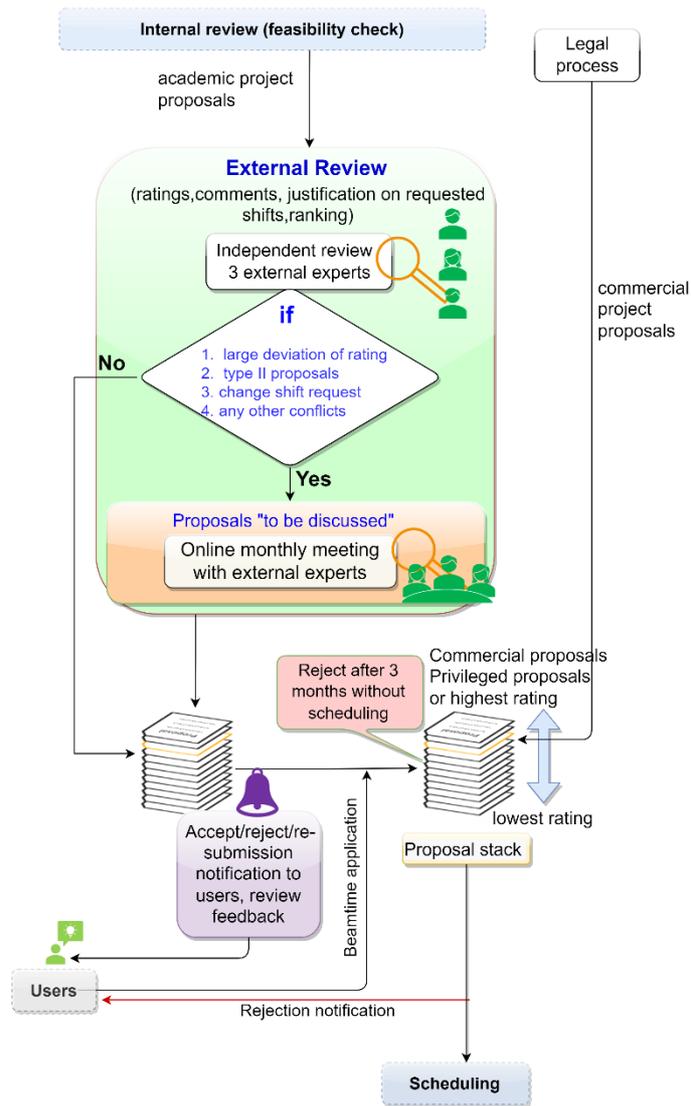

Figure 2. Schematic diagram of the external review procedure for the generic rolling access model. This process combines independent reviews with monthly online meetings including the external experts to finalize ratings, provide comments, establish rankings, and resolve conflicts.

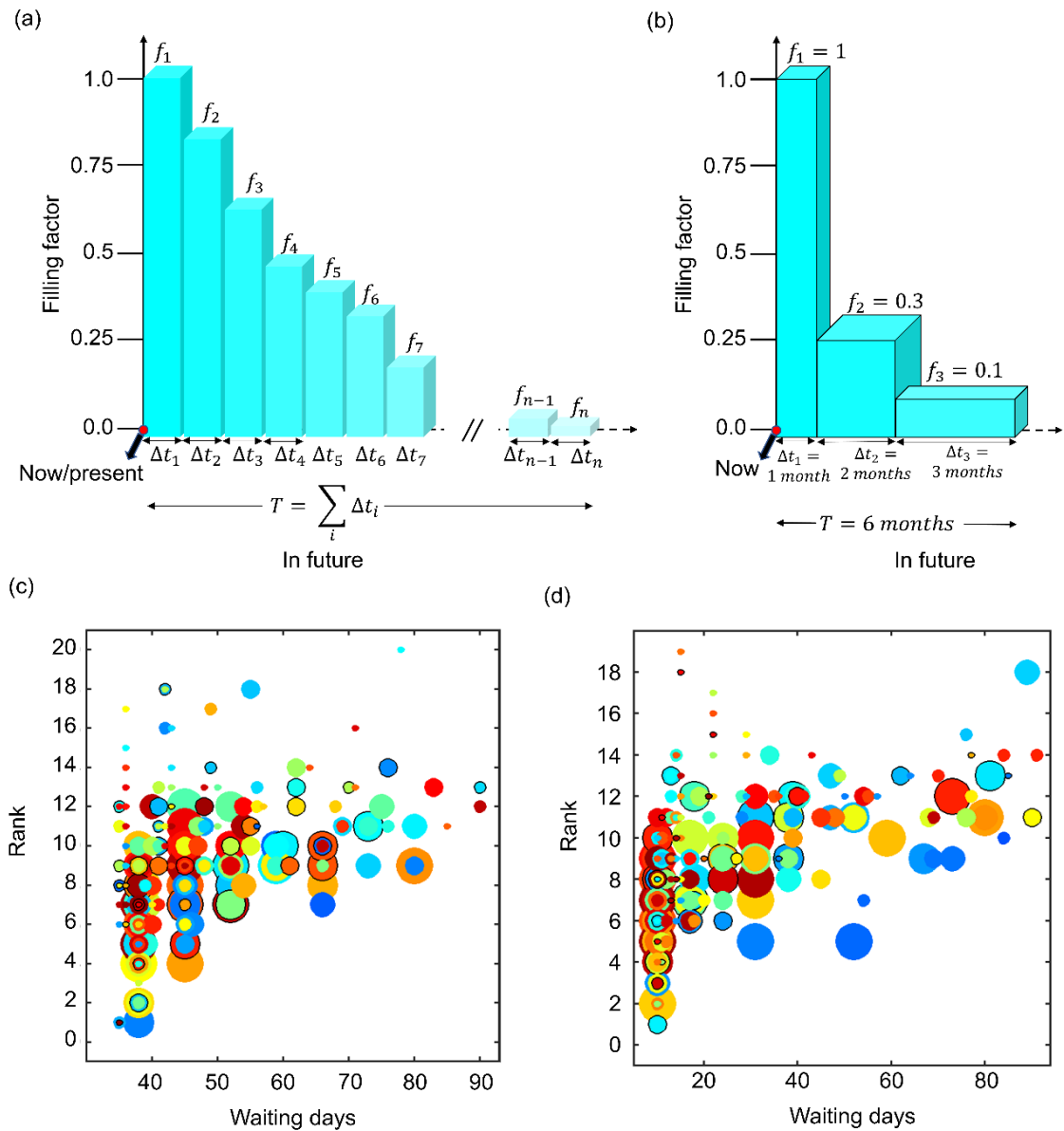

Figure 3. (a) Sketch of the new dynamic scheduling strategy for the generic rolling access model with a scheduling period sliced in n-periods, (b) A specific example with three-period model (case 1) scheduling strategy. (c) and (d) The correlation of the rank of the proposals to the waiting time; (c) for case 1 with parameters ($\Delta t_1$, $\Delta t_2$, $\Delta t_3$, $f_1$, $f_2$, $f_3$, $c$, $u$) = (30 days, 60 days, 90 days, 1, 0.3, 0.1, 3 months, 1 week), (d) case 2 with parameters ($\Delta t_1$, $\Delta t_2$, $\Delta t_3$, $f_1$, $f_2$, $f_3$, $c$, $u$) = (15 days, 85 days, 55 days, 1, 0.15, 0.12, 3 months, 1 week). The simulation has been performed with 110 project proposals. The experiment duration is encoded in the circle diameter; from small circles (3 shifts) continuously to the largest circles (18 shifts).

Acknowledgement:

The PETRA IV project team (A.B.D., K.B., R.K., H.R.) acknowledges funding of the technical design phase granted by the Behörde für Wissenschaft, Forschung, Gleichstellung und Bezirke (BWFGB) of the Freie und Hansestadt Hamburg under contract BWFG/F|97236 and granted by the Bundesministerium für -Forschung, Technologie und Raumfahrt (BMFTR) under the contract DES21TDR. The PETRA IV project team further acknowledges the funding of the preparation phase by the Bundesministerium für -Forschung, Technologie und Raumfahrt (BMFTR) under the contract VPETRAIV.

Declaration of competing interest

The authors declare that they have no known competing financial interests.